\documentclass[12pt]{article}

\usepackage{amsmath}
\usepackage[makeroom]{cancel}
\usepackage{graphicx}
\usepackage{amssymb}
\usepackage{amsfonts}
\usepackage{latexsym}
\usepackage{hyperref}
\usepackage[english]{babel}
\usepackage{framed,color}
\usepackage{enumerate}
\DeclareMathOperator{\cx}{\square}
\def\beq{\begin{eqnarray}}
\def\eeq{\end{eqnarray}}

\def\al{\alpha}
\def\be{\beta}

\def\ga{\gamma}
\def\de{\delta}
\def\vp{\varepsilon}
\def\ep{\epsilon}

\def\ka{\kappa}
\def\la{\lambda}
\def\na{\nabla}

\def\rh{\rho}
\def\si{\sigma}

\def\ta{\tau}

\def\Ga{\Gamma}
\def\De{\Delta}

\def\Si{\Sigma}
\def\Om{\Omega}

\begin{document}

\begin{center}
{\large\bf
On the basis of surface curvature-dependent terms in $6D$}
\vskip 4mm

Fabricio M. Ferreira~$^{a,b}\,$\footnote{
E-mail address: fabricio.ferreira@ifsudestemg.edu.br}
\
and
\
Ilya L. Shapiro~$^{a,c,d}\,$\footnote{
E-mail address: shapiro@fisica.ufjf.br}

%%%%%%%%%%%%%%%%%%%%%%%%%%%%%
\vskip 4mm

{\sl
(a) \ Departamento de F\'{\i}sica, ICE,
Universidade Federal de Juiz de Fora
\\
Campus Universit\'{a}rio - Juiz de Fora, 36036-330, MG, Brazil}
\vskip 2mm

{\sl
(b) \ Instituto Federal de Educa\c c\~ao, Ci\^encia e Tecnologia
do Sudeste de Minas Gerais \\
IF Sudeste MG - Juiz de Fora, 36080-001, MG, Brazil}
\vskip 2mm

{\sl
(c) \ Tomsk State Pedagogical University, 634034, Tomsk, Russia}
\vskip 2mm

{\sl
(d) \
Tomsk State University, 634050, Tomsk, Russia}
\vskip 2mm

\end{center}

\vskip 6mm
%%%%%%%%%%%%%%%%%%%%%%%%%%%%%%%%

\begin{quotation}
\noindent
\textbf{Abstract.} \
Total derivative terms play an important role in the integration of
conformal anomaly. In four dimensional space $4D$ there is only one
such term, namely ${\cx}R$. In the case of six dimensions $6D$
the structure of surface terms is more complicated, and it is useful
to construct a basis of linear independent total derivative terms. We
briefly review the general scheme of integrating anomaly and present
the reduction of the minimal set of the surface terms in $6D$ from
eight to seven. Furthermore, we discuss the comparison with the
previously known equivalent reduction based on the general covariance
and obtain it also from the conformal symmetry. Our results confirms
that the anomaly induced effective action in $6D$ really has a
qualitatively new (compared to previously elaborated $2D$ and
$4D$ cases) ambiguity, which is parameterized by the two parameters
$\xi_1$ and  $\xi_2$.
\vskip 2mm

\noindent
{\it Keywords:} \
Total derivatives terms,  Higher dimensions,  Conformal anomaly
\vskip 2mm

\noindent
{\it PACS:} \
04.62.+v, \  %%	Quantum fields in curved spacetime
04.50.-h  % Higher-dimensional gravity and other theories of gravity
\end{quotation}

%%%%%%%%%%%%%%%%%%%%%%%%%%
%%%%%%%%%%%%%%%%%%%%%%%%%%
%%%%%%%%%%%%%%%%%%%%%%%%%%
\section{Introduction}
\label{intro}

The integration of trace anomaly is the simplest way to derive the
effective action (EA) of vacuum. The anomaly-induced action proved
being a powerful tool due to the compact and useful form of the
result, which finds many applications (see, e.g. \cite{Duff-94} for
the review). The integration of anomaly was originally done in the
two dimensional space $2D$ in the important work of Polyakov
\cite{Polyakov81}. The generalization for  $4D$ was done by
Riegert \cite{rie} and Fradkin and Tseytlin \cite{FrTs84}. There
are interesting general features of anomaly, which can not
be seen in $2D$ and can be merely noticed in  $4D$. The reason
is that in  $4D$ there is only one possible surface term ${\cx}R$
in the anomaly, while in  $2D$ there are no such terms at all.

Things change dramatically in $6D$, where we meet a bunch of the
surface terms, which make integration of anomaly quite a challenging
task. As we have discussed in the previous papers \cite{FST,6d},
the number of possible covariant surface terms with the proper
dimension coming from the derivatives of the metric may be larger
\cite{Bonora}, but it can be reduced to eight \cite{6d}. In the
present contribution we show how this number can be reduced further
to seven terms and discuss the relation of the corresponding identity
to the diffeomorphism invariance from one side and with the conformal
property of the Gauss-Bonnet term in $6D$ from another side. Indeed,
the identity itself has been known previously \cite{Oliva:2010zd}, but
in what follows we present its direct derivation and also show the
relation to conformal transformation of the metric.

The paper is organized as follows. In Sec.~\ref{rev} we briefly
review the general scheme of integrating anomaly in even dimension
and present the result of anomaly integration in $6D$. In
Sec.~\ref{6dbasis} we describe the reduction of the basis of
surface terms. Sec.~\ref{2sides} describes how the main reduction
formula is related to the general covariance and conformal invariance
of the term which is topological invariant in $6D$, and why this
identity is valid beyond this particular dimension. Finally, in
Sec.~\ref{AAB1} we draw our conclusions and discuss the
implications of this work for the integration of anomaly.

%%%%%%%%%%%%%%%%%%%%%%%%%%
%%%%%%%%%%%%%%%%%%%%%%%%%%
%%%%%%%%%%%%%%%%%%%%%%%%%%
\section{Anomaly induced effective action}
\label{rev}

The general structure of conformal anomaly in an arbitrary even
dimension $D=2n$ includes the following three types of terms:

i) Conformal invariant structures, such as
$C_{\mu\nu\al\be}^2=C_{\mu\nu\al\be}C^{\mu\nu\al\be}$ in $4D$.
In the simplest case of $2D$ there are no conformal term, while in higher
dimensions there may be much more such terms $\sum c_r\,W^{r}_D$,
with the sum over $r$. For instance, there are three of them in $6D$
\cite{BFT-2000}.

ii) The topological invariant
\beq
\label{E2n}
E_{(2n)}
\,=\,
\frac{1}{2^{n}}\,
\vp^{\al_{1} \be_{1}\,\dots\,\al_n \be_n}\,
\vp^{\ga_{1} \de_{1}\,\dots\,\ga_n \de_n}
\,R_{\al_1 \be_1 \ga_1 \de_1}\,\,\dots\,\,R_{ \al_n \be_n \ga_n \de_n }.
%% \, , n \in \mathbb{N}^* .
\label{caract}
\eeq

iii) The set of surface terms $\Xi_D=\sum \ga_k\chi_k$. In $4D$ there
is only one surface term $\Box R$, and in higher dimensions there are
always much more such terms. The main subject of the present
communication is the reduction of the minimal set of surface terms
in $6D$.

The reason for the described classification of constituents of the trace
anomaly (see, e.g., \cite{DDI,Duff77,DeserSchwimmer}) is that the
anomaly reflects the form of the one-loop divergences in the vacuum
sector, and the last satisfy conformal Noether identity in case when
quantized matter fields are conformal. Thus the anomaly can be
presented in the universal form
\beq
T \,=\, \langle T_\mu^\mu \rangle
\,=\, c_r\,W^{r}_D \,+\,a\,E_D \,+\,\Xi_D.
\label{T}
\eeq

For the integration of anomaly it proves useful to start from the conformal
transformation of the metric tensor $\,g_{\mu \nu}$,
\beq
\label{tconf01}
g_{\mu \nu}= e^{2\si(x)} \bar{g}_{\mu \nu}\,.
\eeq
The main ingredient of the scheme described in \cite{rie,FrTs84} is the
transformation rule for the corrected topological invariant,
\beq
\label{conjec}
\sqrt{-g}\tilde{E}_{D} \, =\,\sqrt{-\bar{g}}\big(\bar{\tilde{E}}_{D}
+\ka \bar{\De}_{D}\si\big) ,
\eeq
where $D = 2,4,6,\cdots$ and
\beq
\tilde E_D  %% \,=\, E_D + \na_\mu \xi^\mu
\,=\, E_D + \sum_{i}\al_i \Xi_i
\label{ED}
\eeq
is the modified Euler density which is a sum of the original
topological term $E_D$ and a special linear combination of the total
derivatives of the curvature-dependent terms $\Xi_i$ with the
coefficients $\al_i $. Finally, $\De_D$ is the conformal operator,
which is a $D$-dimensional generalization of the Paneitz operator
in $4D$, \cite{FrTs-superconf,Paneitz}. The $6D$ solution for $\al_i$
is \cite{6d}
\beq
\label{coef}
&&
\al_1 = \frac{3}{5}
,\quad
\al_2 = -\frac{9}{10} -\frac{5}{4}\xi_1 +\frac{3}{8} \xi_2
,\quad
\al_3 = \xi_1,
\quad
\al_4 = 0
,\quad
\al_5 = \frac{84}{5} +3\xi_1 + \frac{11}{2}\xi_2,
\nonumber
\\
&&
\al_6 = -\frac{36}{5} - 2\xi_1 -5 \xi_2
,\quad
\al_7 = -\frac{18}{5}-\xi_1-\frac{7}{2}\xi_2
,\quad
\al_8 = \xi_2 .
\mbox{\quad}
\eeq
Here $\xi_1$ and $\xi_2$ are two arbitrary  parameters which can be
fixed only if we find more than one identically vanishing linear
combination of the surface terms. Indeed, the two-parameter ambiguity
in the conformal operator  $\bar{\De}_{D}$ has been found in the
paper \cite{Hamada}.

As far as the coefficients $\al_i$ in Eq. (\ref{ED}) are established,
the  integration of conformal anomaly becomes a relatively simple
exercise, and the general answer can be written in the form \cite{6d}
\beq
\Ga_{ind}
&=&
S_c
+\iint\limits_{x\,y} \Big\{
\frac{1}{4}\,c_r\,W^{r}_D
\,+\,
\frac{a}{8}\,{\tilde E}_D(x)\Big\}
\,G(x,y)\,{\tilde E}_D(y)
\nonumber
\\
&+&   % \limits
\sum_k \big( \ga_k - \al_k \big) \sum_i c_{ik}
\int_x{\cal L}_i\,.
\label{Ga}
\eeq
Here $\,S_c=S_c[g_{\mu\nu}]\,$ is an arbitrary conformal invariant
functional, $\int_x \equiv \int d^Dx \sqrt{-g}$, $\,G(x,y)\,$ is the
Green function of the conformal operator $\De_D$ and, finally,
${\cal L}_i$ are local Lagrangians which generate the surface
terms in the anomaly through the relations,
\beq
-\,\frac{2}{\sqrt{-g}}\,g_{\mu\nu}
\,\frac{\de}{\de g_{\mu\nu}}\,
\sum\limits_i c_{ik} \int_x{\cal L}_i \,=\,\chi_k.
\label{confLoc}
\eeq

One can see that surface terms $\chi_k$ play a decisive role in the
integration of anomaly. Therefore it is very desirable to establish a
minimal set of linear independent surface terms. According to the
previous publications, e.g. \cite{Bonora} or \cite{6d} there are
eight such terms. In the next section we show how this number can
be reduced to seven.

%%%%%%%%%%%%%%%%%%%%%%%%%%%%%%%%%%%%%%
\section{Reduction of six derivative surface terms}
\label{6dbasis}

The set of six derivative surface terms which was used in  \cite{6d}
looks as follows:
\beq
\label{oito}
&&
\Xi_1 = {\cx}^2 R
, \quad
\Xi_2 =
\Box R^2_{\mu \nu \al \be}
, \quad
\Xi_3 =
\Box R^2_{\mu \nu}
, \quad
\Xi_4 = \Box R^2
, \quad
\Xi_5 = \na_\mu \na_\nu \big(
R^{\mu}\,_{\la \al \be} R^{\nu \la \al \be}\big),
\nonumber
\\
&&
\Xi_6= \na_\mu \na_\nu \big(R_{\al \be} R^{\mu \al \nu \be}\big)
,\quad
\Xi_7 = \na_\mu \na_\nu \big(R_{\al}^{\mu} R^{\nu \al}\big)
, \quad
\Xi_8 =  \na_\mu \na_\nu \big(R R^{\mu \nu} \big).
\eeq

Let us start with the following statement which can be obtained by direct
calculation. Performing the conformal transformations of the structures
$\Xi_k$ one can prove, with the help of the software {\sl Mathematica}
\cite{Wolfram}, that the following linear combination of surface terms
is conformal invariant:
\beq
\label{Xi_comb}
\sqrt{-g} \big(
\Xi_2 -4\Xi_3 + \Xi_4
-4\Xi_5+ 8\Xi_6 + 8\Xi_7 -4\Xi_8 \big)
\nonumber
\\
= \sqrt{-\bar{g}}\big(
\bar\Xi_2 -4\bar\Xi_3 + \bar\Xi_4
-4\bar\Xi_5+ 8\bar\Xi_6 + 8\bar\Xi_7 -4\bar\Xi_8
\big).
\eeq

Could the above combination be identically vanishing, indicating
linear dependence of the set (\ref{oito})? The answer to this
question is positive, and we demonstrate this in what follows.
For the sake of this proof, we introduce the following notations:
\begin{align}
\label{base_A}
&\Si_1 = {\cx}^2{R}
&& \Si_2 =  (\na_{\la}{R}_{\mu \nu \al \be})^2
&& \Si_3 = {R}_{\mu \al \nu \be}\na^{\mu}\na^{\nu} {R}^{\al \be}
 \nonumber\\
 &
 \Si_4 = {R}_{\mu \nu} {R}^{\mu \la \al \be} {R}^{\nu}\,_{\la \al \be}
&& \Si_5 = R^{\mu \nu}\,_{\al \be} R^{\al \be}\,_{\la \ta} R^{\la \ta}\,_{\mu \nu}
&& \Si_6 =  R^{\mu}\,_{\al}\,^{\nu}\,_{\be} R^{\al}\,_{\la}\,^{\be}\,_{\ta}
           R^{\la}\,_{\mu}\,^{\ta}\,_{\nu}
 \nonumber\\
 &\Si_7 = (\na_{\la}{R}_{\mu \nu})^2
&& \Si_8 =  {R}_{\mu \nu}{\cx}{R}^{\mu \nu}
&& \Si_9 =  (\na_{\mu}{R})^2
\nonumber\\
&\Si_{10} = {R}{\cx}{R}
&& \Si_{11} =  (\na_{\al}{R}_{\mu \nu})\na^{\mu}{R}^{\nu \al}
&& \Si_{12} =  R^{\mu \nu} \na_{\mu}\na_{\nu} R
\nonumber\\
&\Si_{13} =  R_{\mu \nu} R_{\al \be} {R}^{\mu \al \nu \be}
&& \Si_{14} = R_{\mu \nu} R^{\mu \al}R_{\al}^{\nu}\,.
&& \,
\end{align}

Let us elaborate each of the terms $\Xi_i$ using notations (\ref{base_A}).
We start from the trivial simplest case and then go the more complicated
part.
\beq
\label{rel_01}
&&
{\cx}^2{R} = \Si_1.
\\
&&
{\cx}{R}^2_{\mu \nu \al \be} = 2(\na_{\la}{R}_{\mu \nu \al \be})^2
+ 2{R}_{\mu \nu \al \be}{\cx}{R}^{\mu \nu \al \be} .
\eeq
Using the properties of the Riemann tensor and Bianci identities,
the second term in the last relation can be rewritten as \cite{decanini}
\beq
{R}_{\mu \nu \al \be}{\cx}{R}^{\mu \nu \al \be}
& = &
4 R_{\mu \al \nu \be} \na^{\mu} \na^{\nu}R^{\al \be}
+2 R_{\mu \nu} R^{\mu \la \al \be} R^{\nu}\,_{\la \al \be}
\nonumber
\\
&&
-R^{\mu \nu}\,_{\al \be} R^{\al \be}\,_{\la \ta} R^{\la \ta}\,_{\mu \nu}
-4  R^{\mu}\,_{\al}\,^{\nu}\,_{\be} R^{\al}\,_{\la}\,^{\be}\,_{\ta} R^{\la}
\,_{\mu}\,^{\ta}\,_{\nu}\, .
\eeq
Thus we arrive at the first relations
\beq
\label{rel_02}
&&
{\cx}{R}^2_{\mu \nu \al \be} =
2 \Si_2
+8 \Si_3
+4 \Si_4
-2 \Si_5
-8 \Si_6 ,
\\
&&
{\cx}{R}^2_{\mu \nu} =
 2(\na_{\la}{R}_{\mu \nu})^2 + 2{R}_{\mu \nu}{\cx}{R}^{\mu \nu}
=  2\Si_7 + 2\Si_8,
\label{rel_03}
\\
&&
{\cx}{R}^2 =  2(\na_{\la}{R})^2 + 2{R}{\cx}{R}
=  2\Si_9 + 2\Si_{10}.
\label{rel_04}
\eeq

Furthermore,
\beq
\label{DDRieRie01}
&&
\nabla_{\mu}\nabla_{\nu}
\big(R^{\mu}\,_{\la \al \be} R^{\nu \la \al \be}\big)
\,=\,
\na_{\mu}\big[(\na_{\nu} R^{\mu}\,_{\la \al \be}) R^{\nu \la \al \be}
+ R^{\mu}\,_{\la \al \be} \na_{\nu}  R^{\nu \la \al \be}\big]
\nonumber
\\
&=&
\na_{\mu}\big[(\na_{\nu} R^{\mu}\,_{\la \al \be}) R^{\nu \la \al \be}
+ R^{\mu}\,_{\la \al \be} \na^{\al}  R^{\la \be}
- R^{\mu}\,_{\la \al \be} \na^{\be}  R^{\la \al}\big]
\nonumber
\\
&=&
\big(\na_{\mu} \na_{\nu} R^{\mu}\,_{\la \al \be}) R^{\nu \la \al \be}
+\big( \na_{\nu} R^{\mu}\,_{\la \al \be}) \na_{\mu} R^{\nu \la \al \be}
+ \big( \na_{\mu} R^{\mu}\,_{\la \al \be} \big) \na^{\al}   R^{\la \be}
\nonumber
\\
&+&
R^{\mu}\,_{\la \al \be}  \na_{\mu} \na^{\al}   R^{\la \be}
- \big( \na_{\mu} R^{\mu}\,_{\la \al \be} \big) \na^{\be}  R^{\la \al}
- R^{\mu}\,_{\la \al \be} \na_{\mu} \na^{\be}  R^{\la \al}
\\
&=&
\big(\na_{\mu} \na_{\nu} R^{\mu}\,_{\la \al \be}) R^{\nu \la \al \be}
+\big( \na_{\nu} R^{\mu}\,_{\la \al \be}) \na_{\mu} R^{\nu \la \al \be}
%% \nonumber \\ &&
+ 2\big( \na_{\mu} R^{\mu}\,_{\la \al \be} \big) \na^{\al}   R^{\la \be}
+  2 \Si_3.
\nonumber
\eeq

The first term in the expression (\ref{DDRieRie01}) can be
transformed as follows:
\beq
\label{DDRieRie02}
&& \big(\na_{\mu} \na_{\nu} R^{\mu}\,_{\la \al \be})R^{\nu \la \al \be}
\,=\,
\big(\na_{\nu} \na_{\mu} R^{\mu}\,_{\la \al \be})R^{\nu \la \al \be}
+ R^{\nu \la \al \be} \big[\na_{\mu}, \na_{\nu}\big] R^{\mu}\,_{\la \al \be}
\nonumber
\\
&=&
\big( \na_{\nu} \na_{\al} R_{\la  \be})R^{\nu \la \al \be}
- \big( \na_{\nu} \na_{\be} R_{\la \al})R^{\nu \la \al \be}
+ R_{\ka  \nu} R^{\ka}\,_{\la \al \be} R^{\nu \la \al \be}
\nonumber
\\
& - &
 R^{\ka}\,_{\la \mu \nu} R^{\mu}\,_{\ka \al \be} R^{\nu \la \al \be}
- R^{\ka}\,_{\al \mu \nu} R^{\mu}\,_{\la \ka \be}  R^{\nu \la \al \be}
- R^{\ka}\,_{\be \mu \nu} R^{\mu}\,_{\la \al \ka}  R^{\nu \la \al \be}
\nonumber
\\
&=& 2 \Si_3 + \Si_4 -  R^{\ka}\,_{\la \mu \nu} R^{\mu}\,_{\ka \al \be} R^{\nu \la \al \be}
- 2R^{\ka}\,_{\al \mu \nu} R^{\mu}\,_{\la \ka \be}  R^{\nu \la \al \be}.
\eeq

At this moment we remember that
\beq
R^{\ka}\,_{\la \mu \nu} R^{\mu}\,_{\ka \al \be} R^{\nu \la \al \be}
&=&
R_{\ka \la \mu \nu} R^{\mu \ka}\,_{ \al \be} R^{\al \be \nu \la }
\nonumber
\\
&=&
-R_{\ka \nu \la \mu } R^{\mu \ka}\,_{ \al \be} R^{\al \be \nu \la }
-R_{\ka \mu \nu \la } R^{\mu \ka}\,_{ \al \be} R^{\al \be \nu \la }.
\eeq
By making change of indices $\nu \leftrightarrow \la$ in the first
of these expressions, we arrive at
\beq
R^{\ka}\,_{\la \mu \nu} R^{\mu}\,_{\ka \al \be} R^{\nu \la \al \be}
\,=\,
- R_{\ka  \la \nu \mu } R^{\mu \ka}\,_{ \al \be} R^{\al \be \la \nu }
+ \Si_5
\,=\, \frac12 \Si_5 .
\eeq
Next, the last term of (\ref{DDRieRie02}) can be transformed as
\beq
R^{\ka}\,_{\al \mu \nu} R^{\mu}\,_{\la \ka \be}  R^{\nu \la \al \be}
&=&
R^{\ka}\,_{\al}\,^{\mu}\,_{\nu}
R_{\mu}\,^{\la}\,_{\ka}\,^{\be}  R^{\nu}\,_{ \la}\,^{\al}\,_{\be}
% \nonumber \\ &=&
= R^{\mu}\,_{\nu}\,^{\ka}\,_{\al} R^{\nu}\,_{ \la}\,^{\al}\,_{\be}
R^{\la}\,_{\mu}\,^{\be}\,_{\ka}=\Si_6 ,
\quad
\eeq
thus we get
\beq
\label{Term01}
\big(\na_{\mu} \na_{\nu} R^{\mu}\,_{\la \al \be})R^{\nu \la \al \be}
&=&
2 \Si_3 + \Si_4 - \frac{1}{2}\Si_5 -2 \Si_6 .
\eeq
The second term of the second expression of (\ref{DDRieRie01}) can
be developed as
\beq
\big( \na_{\nu} R_{\mu \la \al \be}) \na^{\mu} R^{\nu \la \al \be}
=- \big( \na_{\la} R_{\nu \mu \al \be}) \na^{\mu} R^{\nu \la \al \be}
- \big( \na_{\mu} R_{\la \nu \al \be}) \na^{\mu} R^{\nu \la \al \be}.
\eeq
By using the index exchange $\nu \leftrightarrow \la$ in the last
formula we get
\beq
\big( \na_{\nu} R_{\mu \la \al \be}) \na^{\mu} R^{\nu \la \al \be}
&=& - \big( \na_{\nu} R_{\la \mu \al \be}) \na^{\mu} R^{\la \nu \al \be}
- \big( \na_{\mu} R_{\la \nu \al \be}) \na^{\mu} R^{\nu \la \al \be}
\nonumber
\\
&=&
- \big( \na_{\nu} R_{\mu \la \al \be}) \na^{\mu} R^{\nu \la \al \be}
+ \Si_{2},
\eeq
hence
\beq
\label{Term02}
\big( \na_{\nu} R_{\mu \la \al \be}) \na^{\mu} R^{\nu \la \al \be}
= \frac{1}{2}\Si_{2}.
\eeq
Using the first reduced Bianchi identity, we develop the third term
of (\ref{DDRieRie01}) such that
\beq \label{Term03}
\big( \na_{\mu} R^{\mu}\,_{\la \al \be} \big) \na^{\al}   R^{\la \be}
&=&
\big( \na_{\al} R_{\la \be} \big) \na^{\al}   R^{\la \be}
- \big( \na_{\be} R_{\la \al} \big) \na^{\al}   R^{\la \be}
\,=\, \Si_7 - \Si_{11}.
\eeq
Replacing (\ref{Term01}), (\ref{Term02}) e (\ref{Term03}) into
(\ref{DDRieRie01}) we obtain
\beq
\label{rel_05}
\nabla_{\mu}\nabla_{\nu}(R^{\mu}\,_{\la \al \be} R^{\nu \la \al \be})
= \frac{1}{2}\Si_{2} + 4\Si_3 + \Si_4 -\frac{1}{2}\Si_5 - 2\Si_6
+ 2\Si_7  - 2\Si_{11}.
\eeq

The next step is to consider
\beq
&&
\na_{\mu}\na_{\nu}(R_{\al \be} R^{\mu \al \nu \be})
\,=\,
\na_{\mu}
\big[ ( \na_{\nu} R_{\al \be} ) R^{\mu \al \nu \be}
+  R_{\al \be}  \na_{\nu} R^{\nu \be \mu \al } \big]
\nonumber
\\
&=&
\na_{\mu}\big[ ( \na_{\nu} R_{\al \be} ) R^{\mu \al \nu \be}
+  R_{\al \be}  \na^{\mu} R^{\be  \al}
-  R_{\al\be}  \na^{\al} R^{\be  \mu} \big]
\nonumber
\\
&=&
(\na_{\mu} \na_{\nu} R_{\al \be} ) R^{\mu \al \nu \be}
+ ( \na_{\nu} R_{\al \be} ) \na_{\mu} R^{\mu \al \nu \be}
+ (\na_{\mu} R_{\al \be})^2
\nonumber
\\
&+&
R_{\al \be} {\cx} R^{\al \be}
- ( \na_{\mu} R_{\al \be} ) \na^{\al} R^{\be  \mu}
-   R^{\al}_{\be}  \na_{\mu}\na_{\al} R^{\be  \mu}
\nonumber
\\
&=&
\Si_3 + ( \na_{\nu} R_{\al \be} )^2
-( \na_{\nu} R_{\al \be} ) \na^{\be} R^{\al \nu}
+ \Si_7 + \Si_8 -\Si_{11}
\nonumber
\\
&-&
\frac{1}{2}R^{\al\be} \na_{\al}\na_{\be}R
- R^{\al}_{\be}  \big[\na_{\mu}, \na_{\al} \big] R^{\be \mu}
\nonumber
\\
&=&
\Si_3  + 2 \Si_7 + \Si_8 -2\Si_{11}  - \frac{1}{2}\Si_{12}
- R^{\al}_{\be} R^{\be}\,_{\ka \mu \al} R^{\ka \mu}
- R^{\al}_{\be} R_{\ka \al} R^{\be \ka},
\eeq
that means
\beq
\label{rel_06}
\nabla_{\mu}\nabla_{\nu}(R_{\al \be} R^{\mu \al \nu \be})
= \Si_3  + 2 \Si_7 + \Si_8 -2\Si_{11} -\frac{1}{2}\Si_{12}
+\Si_{13} - \Si_{14}.
\eeq

Next,
\beq
&&
\nabla_{\mu}\nabla_{\nu}(R_{\al}^{\mu} R^{\nu \al})
\,=\,
\nabla_{\mu} \big[(\na_{\nu} R^{\mu}_{\al})R^{\nu \al}
+\frac{1}{2} R^{\mu}_{\al}\na^{\al}R \big]
\nonumber
\\
&=&
(\nabla_{\mu}\nabla_{\nu} R_{\al}^{\mu}) R^{\nu \al}
+(\na_{\nu}R^{\mu}_{\al})\na_{\mu} R^{\nu \al}
 %% \nonumber\\&&
+\frac{1}{4}(\na_{\al} R)^2
+\frac{1}{2}R^{\mu}_{\al} \na_{\mu} \na^{\al} R
\nonumber\\
&=&
\frac{1}{2}( \na_{\nu} \na_{\al} R) R^{\nu \al}
+ \Big( \big[\nabla_{\mu},\nabla_{\nu} \big]  R_{\al}^{\mu}\Big) R^{\nu \al}
+ \frac{1}{4}\Si_9
+ \Si_{11}
+ \frac{1}{2}\Si_{12}
\nonumber
\\
&=&
\frac{1}{4}\Si_9 +\Si_{11} + \Si_{12}
+ R_{\ka \nu} R^{\ka \al} R^{\nu}_{\al}
+R^{\al}\,_{\ka \mu \nu} R^{\mu \ka} R^{\nu}_{\al}.
\nonumber\\
&=&
\frac{1}{4}\Si_9 +\Si_{11} + \Si_{12} -\Si_{13} +\Si_{14}.
\label{rel_07}
\eeq

Finally, simpler operations provide the last ingredients,
\beq
&&
\nabla_{\mu}\nabla_{\nu}(R R^{\mu \nu})
= \frac{1}{2} R{\cx}R
+ (\na_{\mu} R)^2
+ R^{\mu \nu} \na_{\mu}\na_{\nu} R,
\\
&&
\nabla_{\mu}\nabla_{\nu}(R R^{\mu \nu})
= \Si_{9} + \frac{1}{2} \Si_{10} +\Si_{12}.
\label{rel_08}
\eeq

Now we possess all what is needed to solve the equation of our
interest,
\beq
a\Xi_1+ b\Xi_2 +c\Xi_3 + d\Xi_4
+e\Xi_5+ f\Xi_6 +g\Xi_7 + h\Xi_8
\equiv 0.
\eeq
The solution for the coefficients of this equation is as follows:
\begin{align}
\label{dep_lin01}
& a = 0
&& b = \be
&& c = -4\be
&& d = \be
\nonumber\\
& e = -4\be && f = 8\be && g = 8\be && h = -4\be ,
\end{align}
where $\be$ is an arbitrary number which can be equal to one.
Therefore, we have proved the identity
\beq
\label{dep_lin}
\Xi_2 -4\Xi_3 + \Xi_4 -4\Xi_5+ 8\Xi_6 + 8\Xi_7 -4\Xi_8 = 0.
\eeq
Eq.~(\ref{dep_lin}) resolves the main problem which we posed
at the beginning of this contribution. Namely, it reduce the number
of linearly independent six-derivative surface terms from eight to
seven. Still this is not a complete solution of all relevant issues
which one meets in the part of surface terms, and one can find the
description of remaining problems in the next section.

%%%%%%%%%%%%%%%%%%%%%%%%%%%%%%%
\section{Two sides of the identity (\ref{dep_lin})}
\label{2sides}

After sending the first version of this manuscript to arXiv we learned
about the well known paper \cite{Oliva:2010zd}, where the identity
(\ref{dep_lin}) has been used for deriving other relations between the
equations of motion of the six-derivative actions in $6D$. The way
this identity has been obtained in the mentioned work came from the
similar consideration in \cite{BD85} for the Gauss-Bonnet invariant
in $4D$. The relation can be obtained as a Noether identity for the
diffeomorphism invariance of the corresponding topological action
(in some formulas we avoid using condensed notation for a $D$
dimensional integral, just to stress its dimension),
\beq
S_{GB}^{(D)}
&=&
\int d^Dx \sqrt{-g} E_D .
\label{GBaction}
\eeq

The Gauss-Bonnet term (\ref{caract}) in $6D$ can be cast into
the form
\beq
E_6
&=&
\frac{1}{8}\,
\vp^{\al_1\be_1\al_2 \be_2\al_3 \be_3}\,
\vp^{\ga_1 \de_1\,\dots\,\ga_3 \de_3}
\,R_{\al_1 \be_1 \ga_1 \de_1}\,\,\dots\,\,R_{ \al_3 \be_3 \ga_3 \de_3 }
\nonumber
\\
&=&
-8{\cal L}_1
+4{\cal L}_2
-24{\cal L}_3
+24{\cal L}_4
+16{\cal L}_5
+3{\cal L}_6
-12{\cal L}_7
+{\cal L}_8.
\label{carac6}
\eeq
It is interesting that the first of these presentations does not admit
simple generalization to an arbitrary dimension $D$, while for the
second one it is not an obstacle. In what follows we will assume
that $E_6$ means the expression in the {\it r.h.s.} when it is
considered in $D \neq 6$.

The Noether identity for the general covariance of the action has the form
\beq
\na_\mu \bigg[
\frac{2}{\sqrt{-g}}\,
\frac{\de}{\de g_{\mu\nu}}\,S_{GB}^{(D)}
\bigg]
&=&
0.
\label{GBeqn}
\eeq
The last identity reflects only the covariance of the action
(\ref{GBaction}) and does not not use the topological nature
of this expression. Therefore this identity is going to hold even
for the dimension $D$ where this action is not topological. At the
same time, since in the ``proper'' dimension the equation of motion
for the topological action is supposed to vanish (see \cite{capkim}
and the book \cite{Tensors} for detailed discussion), its trace is
also vanishing \cite{FST}. One can anticipate that in the case of the
action  (\ref{GBaction}) this can produce another identity, which
can be related to (\ref{GBeqn}) due to the topological nature of
the action in  in the ``proper'' dimension. Let us check the situation
in the case of $6D$.

In what follows we will need the list of the six-derivative actions which
are not full derivatives. One can define these actions in the form
$I_n = \int_x {\cal L}_n$, where
\beq
&&
{\cal L}_1 =
R^\al\,_\la\,^\be\,_\tau
R^\la\,_\rh\,^\tau\,_\si
R^\rh\,_\al\,^\si\,_\be
, \quad
{\cal L}_2 =
R^{\al \be}\,_{\la \tau}
R^{\la \tau}\,_{\rh \si}
R^{\rh \si}\,_{\al \be}
, \quad
{\cal L}_3 =
R_{\al \be}
R^{\al}\,_{\ga \la \tau}
R^{\be \ga \la \tau},
\nonumber
\\
&&
{\cal L}_4=
R^{\al \be}
R^{\la \tau}
R_{\al \la \be \tau}
, \quad
{\cal L}_5 =
R^\al_\la
R^\be_\al
R^\la_\be
, \quad
{\cal L}_6 =
R R^2_{\al \be \la \tau}
, \quad
{\cal L}_7 = R R^2_{\al \be}
,
\nonumber
\\
&&
{\cal L}_8 = R^3
, \quad
{\cal L}_9 =R^{\al \be}{\cx}R_{\al \be}
, \quad
{\cal L}_{10} = R{\cx}R.
\eeq
Furthermore, let us give the list of the corresponding equations of
motion \cite{decanini} (see also \cite{LPP13,Oliva:2010zd})
\beq
\Phi^{\mu \nu}_n
&=&
\frac{1}{\sqrt{-g}}\frac{\de I_n}{\de g_{\mu \nu}}
\quad
\mbox{and their traces}
\quad
\Phi_n
\,\,=\,\,
g_{\mu \nu}\, \Phi^{\mu \nu}_n,
\eeq
which have the following form:
\beq
\Phi^{\mu \nu}_1
&=&
\frac12 g^{\mu \nu}
		R^\al\,_\la\,^\be\,_\tau
		R^\la\,_\rh\,^\tau\,_\si
		R^\rh\,_\al\,^\si\,_\be
	-3R^\al\,_\la\,^{(\mu}\,_\tau
	  R_{\rh \al \si}\,^{\nu)}
	  R^{\la \rh \ta \si}
\nonumber
\\
&&
	+3\na_\la \na_\be
	( R^\la\,_\rh\,^{(\mu}\,_\si
  	R^{\nu) \rh \be \si})
	-3\na^\la \na^\tau
	(R^{(\mu}\,_ \rh \,^{\nu)}\,_\si
	R^\rh\,_\la\,^\si\,_\ta ),
\nonumber
\\
\Phi_1 &=&
\frac{D-6}{2}R^\al\,_\la\,^\be\,_\tau
			R^\la\,_\rh\,^\tau\,_\si
			R^\rh\,_\al\,^{\si}\,_\be
 			+\frac32 \Xi_5 -  3\Xi_6,
\label{phi01a}
\eeq
\beq
\Phi^{\mu \nu}_2
&=&
\frac12 g^{\mu \nu}
R^{\al \be}\,_{\la \tau}
R^{\la \tau}\,_{\rh \si}
R^{\rh \si}\,_{\al \be}
-3R^{\al (\mu}\,_{\la \ta}
  R_{\rh \si \al}\,^{\nu)}
  R^{\la \ta \rh \si}
%% \nonumber\\&&
-6\na^\be \na_\ta
(R^{\ta (\mu}\,_{\rh \si}
R_\be\,^{\nu) \rh \si}),
\nonumber
\\
\Phi_2 &=&
\frac{D-6}{2}
		R^{\al \be}\,_{\la \tau}
		R^{\la \tau}\,_{\rh \si}
		R^{\rh \si}\,_{\al \be}
		-6 \Xi_5,
\label{phi02a}
\eeq
%%%%%%%%%%%%%%%%%%  3
\beq
\Phi^{\mu \nu}_3
&=&
\frac12 g^{\mu \nu}
	R_{\al \be}
	R^\al \,_{\ga \la \tau}
	R^{\be \ga \la \tau}
	- R^{(\mu}_{\al}
	  R^{\nu)\ga \la \tau}
	  R^\al\,_{\ga \la \tau}
    -2 R_{\al\be} R^{\al}\,_\ga\,^{(\mu}\,_{\tau}
R^{\nu) \ta \be \ga}
\nonumber
\\
&&
- \frac12 g_{\mu \nu}
\na_\al\na_\be
( R^{\al \ga \la \tau}
R^\be\,_{\ga \la \tau} )
+ \na_\al \na^{(\mu}( R^{ \nu) \ga \la \tau}R^\al\,_{\ga \la \tau})
\nonumber
\\
&&
- \frac12 \cx ( R^\mu\,_{\ga \la \tau} R^{\nu \ga \la \tau})
-2 \na_\ga\na_{\la} (R^{(\mu}_{\al}
R^{\nu) \la \al \ga})
-2 \na_\al \na^\ta(R^\al_\be R^{\be(\mu}\,_\ta\,^{\nu)}),
\nonumber
\\
\Phi_3 &=&
\frac{D-6}{2}R_{\al \be}R^\al\,_{\ga \la \tau} R^{\be \ga \la \tau}
-\frac12 \Xi_2 -\frac{D-2}{2}\Xi_5 -2\Xi_6 - 2\Xi_7\, ,
\label{phi03a}
\eeq
%%%%%%%%%%%%%%%%  4
\beq
\Phi^{\mu \nu}_4
&=&
\frac12 g^{\mu \nu}
R^{\al \be} R^{\la \tau} R_{\al \la \be \tau}
-3  ( R_{ \la \al}\,^{(\mu}\,_\be R^{\nu) \la} R^{\al \be})
	- {\cx} ( R^{\al \mu \be \nu} R_{\al \be})
- \na_\al\na_\be (R^{\mu \nu}R^{\al \be})
\nonumber
\\
&&
- g^{\mu \nu} \na^\al \na^\be
(R^{\la \tau} R_{\la \al  \tau \be} )
+2\na_\la \na^{(\mu} ( R^{\nu) \al \la \be} R_{\al \be} )
+\na_\al\na_\be (R^{\al (\mu} R^{\nu) \be}),
\nonumber
\\
\Phi_4
&=&
\frac{D-6}{2}R^{\al \be} R^{\la \tau} R_{\al \la \be \tau}
-\Xi_3 -(D-2)\Xi_6
+\Xi_7- \Xi_8\, ,
\label{phi04a}
\eeq
%%%%%%%%%%%%%%%%  5
\beq
\Phi^{\mu \nu}_5
&=&
\frac12 g^{\mu \nu}  R^{\al}_{\la} R^{\be}_{\al} R^{\la}_{\be}
-3 R^\mu_\be R^{\nu \la} R^\be_\la
+3\na^\al \na^{(\mu}(R^{\nu)}_\la R^\la_\al  )
\nonumber\\
&&
-\frac32 g^{\mu \nu} \na_\al \na_\be (R^{\la \al} R^\be_\la )
-\frac32 \cx (R^\mu_\la R^{\nu \la} ),
\nonumber
\\
\Phi_5
&=&
\frac{D-6}{2}R^\al_\la R^\be_\al R^\la_\be
-\frac32 \Xi_3 -3\frac{D-2}{2}\Xi_7\, ,
\label{phi05a}
\eeq
%%%%%%%%%%%%%%%%  6
\beq
\Phi^{\mu \nu}_6
&=&
\frac12 g^{\mu \nu}  R R^2_{\al \be \la \tau}
- R^{\mu \nu} R^2_{\al \be \la \tau}
-2R R_{\la \tau}\,^{\al (\mu} R_\al\,^{\nu)\la \tau}
\nonumber\\
&&
+\na^\mu \na^\nu R^2_{\al \be \la \tau}
-g^{\mu \nu } \cx R^2_{\al \be \la \tau}
-4 \na^\be \na^\la (R R_\be\,^{(\mu}\,_\la\,^{\nu)}),
\nonumber
\\
\Phi_6
&=&
	\frac{D-6}{2}R R^2_{\al \be \la \tau}
	-(D-1)\Xi_2 -4 \Xi_8 \, ,
\label{phi06a}
\eeq
%%%%%%%%%%%%%%%%  7
\beq
\Phi^{\mu \nu}_7
&=&
\frac12 g^{\mu \nu} R R^2_{\al \be}
- R^{\mu \nu} R^2_{\al \be}
-2R R^\mu_\la R^{\nu \la}
+\na^\mu \na^\nu (R^2_{\al \be})
- g^{\mu \nu}{\cx} R^2_{\al \be}
\nonumber\\
&&
+2 \na_\al \na^{(\mu}(R ^{\nu) \al} R)
-g^{\mu \nu} \na_\al \na_\be (R R^{\al \be})
-{\cx}( R R^{\mu \nu} ),
\nonumber
\\
\Phi_7
&=&
\frac{D-6}{2}R R^2_{\al \be}
-(D-1)\Xi_3
-\Xi_4 -(D-2)\Xi_8\,,
\label{phi07a}
\eeq
%%%%%%%%%%%%%%%%  8
\beq
\Phi^{\mu \nu}_8
&=&
\frac12 g^{\mu \nu}  R^3
-3 R^{\mu \nu} R^2
+3 \na^\mu \na^\nu R^2
-3 g^{\mu \nu}{\cx} R^2,
\nonumber
\\
\Phi_8
&=&
\frac{D-6}{2} R^3
- 3(D-1) \Xi_4\, ,
\label{phi08a}
\eeq
%%%%%%%%%%%%%%%%  9
\beq
\Phi^{\mu \nu}_9
&=&
\frac12 g^{\mu \nu} R^{\al \be}{\cx}R_{\al \be}
-R^{\al \be} \na^{(\mu} \na^{\nu)}  R_{\al \be}
-2R^{(\mu}_\al {\cx}R^{\nu) \al}
+2 \na^\al \na^{(\mu}{\cx} R^{\nu)}_\al
\nonumber\\
&&
-g^{\mu \nu} \na_\al \na_\be {\cx}R^{\al \be}
-{\cx}^2 R^{\mu \nu}
+2\na^\al (R_{\al \be}\na^{(\mu} R^{\nu) \be} )
-2\na^\al (R^{\be (\mu} \na^{\nu)} R_{\al \be} )
\nonumber\\
&&
+\na^{(\mu} ( R_{\al \be} \na^{\nu)}R^{\al \be})
-\frac12 g^{\mu \nu} \na^\la (R_{\al \be} \na_\la R^{\al \be}),
\nonumber
\\
\Phi_9
&=&
\frac{D-6}{2} R^{\al \be}{\cx}R_{\al \be}
 -\frac{D}2 \Xi_1 -\frac{D}2\Xi_3 +\frac{D-4}{4}\Xi_4
 \nonumber\\
 &&
 +2(D-2)\Xi_6 -2\Xi_7-(D-4)\Xi_8\, ,
\label{phi09a}
\eeq
%%%%%%%%%%%%%%%%  10
\beq
\Phi^{\mu \nu}_{10}
&=&
-\frac12 g^{\mu \nu} (\na_\al R)^2
+(\na^\mu R)(\na^\nu R)
+2\na^\mu \na^\nu {\cx} R
% \nonumber \\ &&
-2g^{\mu \nu} {\cx}^2 R
-2 R^{\mu \nu} {\cx} R,
\nonumber
\\
\Phi_{10} &=&
-\frac{D-6}2 R{\cx}R
-2(D-1) \Xi_1
-\frac{D-2}4 \Xi_4 \, .
\label{phi10a}
\eeq
In the derivation of traces we used expressions given in Apendix
\ref{Aa}.

Taking the last observation and new notation into account, by
combining equations (\ref{phi01a}) - (\ref{phi08a}) we arrive at
the following relation:
\beq
\label{trE6_D}
&&
\frac{1}{\sqrt{-g}}\,g_{\mu \nu}\,\frac{\de}{\de g_{\mu \nu}}
 \int \limits_x  E_6
\,=\,
\frac{D-6}{2} E_6
\nonumber \\ &-&
3(D-5)
\big(
\Xi_2 -4\Xi_3 + \Xi_4
-4\Xi_5+ 8\Xi_6 + 8\Xi_7 -4\Xi_8 \big).
\eeq

The first term in the {\it r.h.s.} of the Eq.~(\ref{trE6_D})
obviously
vanish in $D=6$. At the same time, the {\it l.h.s.} also vanish in
 $D=6$, because in this specific dimension it is the trace of the
 variational derivative of the topological term\footnote{One can
 prove this even without taking trace (see, e.g., \cite{Tensors}),
 but such a proof requires choosing a special coordinate
 system. In general coordinates this equation does not look
 trivial, as it was discussed in \cite{capkim,Gauss}.}. In this sense
 the relation (\ref{trE6_D}) proves that the remaining term in the
 {\it r.h.s.} also vanish in $D=6$. However, this term is exactly an
 identity (\ref{dep_lin}) which we proved directly in the previous
section. It is worth noticing that the proof which we presented
there does not depend on the dimension.

Taking the identity  (\ref{dep_lin}) into account, we arrive at the
simple rule of conformal shift of the term under consideration,
namely
\beq
\frac{g_{\mu \nu}}{\sqrt{-g}}\frac{\de}{\de g_{\mu \nu}}
\int \limits_x  E_6
&=& \frac{D-6}{2} E_6,
\label{shift c}
\eeq
which is perfectly consistent to the main relation of integrating
anomaly (\ref{ED}).

To close the story, let us mention that there is yet another
equivalent form of our identity (\ref{dep_lin})
\beq
\label{traco_gen_del}
\Xi_2 -4\Xi_3 + \Xi_4
- 4 \Xi_5+ 8\Xi_6 + 8\Xi_7 -4\Xi_8
&=&
\frac14 \de^{ \mu \al \be \la \tau} _{\nu \xi \eta \ka \chi}
\na_{\mu} \na^{\nu}
\big(R^{\xi \eta}\,_{\al \be} R^{ \ka \chi}\,_{\la \tau}\big),
\eeq
where (in Euclidean signature)
\beq
\de^{ \mu \al \be \la \tau} _{\nu \xi \eta \ka \chi}
&=&
\ep^{\rho\mu \al \be \la \tau} \ep_{\rho\nu \xi \eta \ka \chi}
\,\,=\,\,
5!\,
\de^{[\mu}_\nu\,
\de^\al_\nu\,
\de^\be_\eta\,
\de^\la_\ka\,
\de^{\tau]}_\chi \,.
\label{delta5}
\eeq

The proof of the relation
\beq
\de^{ \mu \al \be \la \tau} _{\nu \xi \eta \ka \chi}
 \na_{\mu} \na^{\nu}(R^{\xi \eta}\,_{\al \be} R^{ \ka \chi}\,_{\la \tau} )
 \equiv 0
\label{fatal}
\eeq
can be found in Appendix \ref{Bb}.
%%%%%%%%%%%%%%%%%%%%%

%%%%%%%%%%%%%%%%%%%%%%%%%%%%%%
\section{Conclusions and Discussions}
\label{AAB1}

As we have just mentioned above, Eq.~(\ref{dep_lin}) reduce the
number of surface terms which is needed to construct the full
basis of such terms in $6D$. Let us discuss the importance of this
formula in the general context.

The complete and consistent integration of trace anomaly in $6D$
requires several mathematical results which are all not very easy to
accomplish, mainly because the practical calculations in $6D$ are
essentially more involved than the ones in $4D$. At the first place
one needs the main formula (\ref{ED}) which immediately produce
the non-local part of the anomaly-induced action \cite{DDI,6d}. The
general formal expression for this action (\ref{Ga}) for an arbitrary
even dimension has been constructed  in Ref.~\cite{6d}, where we
also reported on the explicit
realization of the key formula (\ref{ED}) in the case of $6D$.
Then, looking at the general expression (\ref{Ga}) we can see
that the remaining part of the effective action is related to the
integration of total derivative terms.

Usually the importance of total derivative terms in the anomaly is
underestimated, since it is supposed that they can be modified or
eliminated by adding finite local counterterms. Such an addition
is a mathematically legal procedure, because the gravitational
vacuum action is not quantized in the framework of semiclassical
theory. Therefore, even if the local nonconformal terms are not
needed for renormalization, one can add them without changing
the general structure of quantum theory in curved space.

In some cases such an addition can be pretty well justified. The main
example of this sort is the Starobinsky inflation \cite{star,star83}
where the $R^2$ term with a very large coefficient is required to
provide the
control over perturbations and, in general, correspondence with the
existing observational data. The attempts to explain the magnitude
of this coefficient from quantum field theory arguments are currently
at the rudimentary level  (see, e.g., \cite{StabInstab}) and hence
the introduction of the large coefficient of $R^2$ is a
phenomenological operation. In general, and especially in $6D$,
there are no observational evidence which can be used to fix the
coefficients of the total derivative terms. Therefore for us the
importance of these terms is certain and without
doubts\footnote{Further
arguments concerning the ambiguities related to local terms in the
induced action can be found in Ref.~\cite{anom2003}, where one
can see also the relation to the non-local structures in the case of
almost vanishing masses of quantum fields.}.

In this situation the formula defining the part of effective action
which comes from the total derivative terms in the anomaly is
(\ref{confLoc}). Then the reduction of the number of the total
derivatives $\chi_k$ in the {\it r.h.s.} of this equation from
eight to seven increases our chances to find the solution. And,
from the general perspective, it would be interesting to have an
independent, new and nontrivial confirmation of the possibility
to integrate total derivatives with the local gravitational terms,
according to Eq.~(\ref{ED}).

Two concluding observations are in order. First of all, since in $6D$
the structure (\ref{GBaction}) is topological, its variational derivative
is zero. At the same time, in $6D$ even the identity for the trace is not
easy to prove explicitly, as the reader could ensure from
Sec.~\ref{6dbasis}. The second aspect is that the topological
term (\ref{caract}) is unique and, therefore, the vanishing linear
combination (\ref{dep_lin}) is also unique\footnote{We are
grateful to Dr. Sourya Ray for stressing this point to us.}  The
important consequence of this uniqueness is that further reduction
of the solution (\ref{coef}) is impossible, because (\ref{caract}) was
already taken into account in \cite{6d}. Thus the main result of the
present work is that now we can affirm that the
fundamental difference between the $2D$ and  $4D$  formulas
(\ref{conjec}) from one side and similar formula in $6D$ from
another side is that in the last case this important formula has
two-parameter ambiguity. The changes of $\xi_1$ or  $\xi_2$ do not
produce a change of conformal functional $S_c$, which is the unique
ambiguous part of the effective action in  $2D$ and  $4D$ cases.
Therefore, now we can claim that in  $6D$ we meet a qualitatively
new kind of ambiguity,  that is something which does not take
place in $2D$ and  $4D$ cases.

% \section{a}
%% \appendix
%
\section{Appendix A. Useful relations for total derivatives.}
\label{Aa}

Let us give useful list of relations for total derivatives,
\beq
\label{deriv001a}
&&
\na_{\mu}\na_{\nu}
\big(R^{\mu}\,_{\al \be \la} R^{\nu \la \be \al}\big) =
\frac12 \Xi_5,
\\
\label{deriv002}
&&
\nabla_{\mu}(R^{\mu\nu}\nabla_{\nu}R)
%% = -\frac{1}{4}{\cx} R^2
%% +\nabla_{\mu}\nabla_{\nu}(RR^{\mu\nu})
=  -\frac14 \Xi_4 +\Xi_8,
\\
\label{deriv003}
&&
\na_\al(R_{\mu \nu} \na^\al R^{\mu \nu}) = \frac12 \Xi_3,
\\
\label{deriv004}
&&
\nabla_{\mu}(R^{\nu}_{\la}\nabla_{\nu}R^{\mu\la})
%% = \nabla_{\mu}\nabla_{\nu}(R^{\mu}_{\la}R^{\nu\la})
%% -\frac{1}{2}\nabla_{\mu}(R^{\mu\nu}\nabla_{\nu}R)
%% \nonumber \\ &=&
= \frac{1}{8}\Xi_4 + \Xi_7 -\frac{1}{2}\Xi_8,
\\
\label{deriv005}
&&
\na_\mu(R^{\mu \al \nu \be} \na_\nu R_{\al \be})
=
%% \na_\mu \na_\nu (R^{\mu \al \nu \be}R_{\al \be})
%% - \frac12 {\cx}R^2_{\mu \nu}
%% +\frac18{\cx}R^2
%% +\na_\mu \na_\nu (R^{\mu}_{\al} R^{\nu \al})
%% -\frac12 \na_\mu \na_\nu (R R^{\mu \nu} )
%% \nonumber \\ &=&
- \frac12\Xi_3
+\frac18\Xi_4
+\Xi_6 +\Xi_7
-\frac12 \Xi_8,
\\
\label{deriv006}
&&
\na_\mu \na_\nu {\cx}R^{\mu \nu}
=
%% \frac12 {\cx}^2 R + \frac12 {\cx} R^2_{\mu \nu}
%% -2\na_{\mu} \na_{\nu} (R^{\mu \al \nu \be} R_{\al \be})
%%  -\frac14 {\cx} R^2 +\na_\mu \na_\nu (R R^{\mu \nu} )
%% \nonumber \\ &=&
\frac12 \Xi_1
+\frac12\Xi_3
-\frac14\Xi_4
-2\Xi_6
+\Xi_8.
\eeq
%

%%%%%%%%%%%%%%%%%% %%%%%%%%%%%%%%%%%%
%%%%%%%%%%%%%%%%%% %%%%%%%%%%%%%%%%%%
%%%%%%%%%%%%%%%%%% %%%%%%%%%%%%%%%%%%
\section{Appendix B. Proof of the relation (\ref{fatal})}
\label{Bb}

Let us denote the object of our interest $\Om$ and take one of the
derivatives,
\beq
 \Om
 &=&
 \de^{ \mu \al \be \la \tau}_{\nu \xi \eta \ka \chi}\,
 \na_{\mu} \na^{\nu}
\big(R^{\xi \eta}\,_{\al \be} R^{ \ka \chi}\,_{\la \tau} \big)
\label{Om}
\nonumber
\\
&=&
 \de^{ \mu \al \be \la \tau}_{\nu \xi \eta \ka \chi}\,\,
\na_{\mu}
 \big[ R^{ \ka \chi}\,_{\la \tau}
 \na^{\nu} R^{\xi \eta}\,_{\al \be}
 +  R^{\xi \eta}\,_{\al \be}
 \na^{\nu} R^{ \ka \chi}\,_{\la \tau}
\big].
\eeq
Using antisymmetry of the object (\ref{delta5}) and the Bianchi
identity, the last expression transforms into
\beq
\label{om01}
\Om
&=&
2\na_{\mu}\big(
\de^{ \mu \al \be \la \tau}_{\nu \xi \eta \ka \chi}
R^{ \ka \chi}\,_{\la \tau}
\na^{\nu} R^{\xi \eta}\,_{\al \be}\big)
\nonumber
\\
&=&
-2\na_{\mu}\Big[
 \de^{ \mu \al \be \la \tau}_{\nu \xi \eta \ka \chi}
 \big( R^{ \ka \chi}\,_{\la \tau}
 \na^{\xi} R^{ \eta \nu}\,_{\al \be}
 + R^{ \ka \chi}\,_{\la \tau}
 \na^{\eta} R^{ \nu \xi}\,_{\al \be}
 \big) \Big].
\eeq
Once again using antisymmetry of (\ref{delta5}) we arrive at
\beq
\label{om02}
 \Om
 &=&
- \,4\na_{\mu}\big( \de^{ \mu \al \be \la \tau}_{\nu \xi \eta \ka \chi}
  R^{ \ka \chi}\,_{\la \tau} \na^{\nu} R^{\xi \eta}\,_{\al \be}\big).
\eeq
Comparing (\ref{om01}) and (\ref{om02}) one can check that
\beq
\Om \,=\, -\,2 \Om \,,
\eeq
which is equivalent to Eq.~(\ref{fatal}).

Let us stress that the analog of this result can be found in \cite{BD85}
for $4D$ and can be also found in \cite{Oliva:2010zd} for $6D$. The
derivation of this identity in both cases was based on the relation
(\ref{GBeqn}) which reflects diffeomorphism invariance of the action
(\ref{GBaction}) with $D=6$ and $E_6$ defined as in the {\it r.h.s.}
of Eq.~(\ref{carac6}). For this reason the identity is valid in any
dimension $D$. At the same time the same identity can be also
obtained in exactly $D=6$ as a Noether identity of the conformal
symmetry (\ref{shift c}).

%%%%%%%%%%%%%%%%%%%%%%%%%%%
%%%%%%%%%%%%%%%%%%%%%%%%%%%
%%%%%%%%%%%%%%%%%%%%%%%%%%%
\section*{Acknowledgements}

Authors are grateful to S.~Ray for stimulating correspondence.
I. Sh. is gratefully acknowledging partial support from CNPq 
(project 303893/2014-1) and FAPEMIG (project APQ-01205-16).

%%%%%%%%%%%%%%%%%%%%%%%%%%%
%%%%%%%%%%%%%%%%%%%%%%%%%%%
%%%%%%%%%%%%%%%%%%%%%%%%%%%


\begin{thebibliography}{m}

\bibitem{Duff-94} M.J. Duff,
{\it Twenty years of the Weyl anomaly,}
Class. Quant. Grav. {\bf 11} (1994) 1387, hep-th/9308075.

\bibitem{Polyakov81} A.M. Polyakov,
{\it Quantum geometry of bosonic strings,}
Phys. Lett. {\bf B103} (1981) 207. %% -210

\bibitem{rie} R.J. Riegert,
{\it A Nonlocal Action for the Trace Anomaly,}
Phys. Lett. {\bf 134B} (1984) 56.

\bibitem{FrTs84} E.S. Fradkin and A.A. Tseytlin,
{\it Conformal Anomaly in Weyl Theory and Anomaly Free Superconformal
Theories,}
Phys. Lett. {\bf 134B} (1984) 187.

\bibitem{FST} F.M. Ferreira, I.L. Shapiro and P.M. Teixeira,
{\it On the conformal properties of topological terms in even
dimensions}
Eur. Phys. J. Plus {\bf 131} (2016) 164,
%% DOI: 10.1140/epjp/i2016-16164-9
arXiv:1507.03620.

\bibitem{6d} F.M. Ferreira, I.L. Shapiro,
{\it Integration of trace anomaly in $6D$,}
Phys. Lett. {\bf B772} (2017) 174. % -178.

\bibitem{Bonora} L. Bonora, P. Pasti and M. Bregola,
{\it Weyl cocycles,}
Class. Quant. Grav. {\bf 3} (1986) 635.

\bibitem{Oliva:2010zd}  J.~Oliva and S.~Ray,
{\it Classification of Six Derivative Lagrangians of Gravity and Static
Spherically Symmetric Solutions,}
Phys. Rev. {\bf D82} (2010) 124030,
%%  doi:10.1103/PhysRevD.82.124030
arXiv:1004.0737.

\bibitem{BFT-2000} 	
F. Bastianelli, S. Frolov, and A.A. Tseytlin,
{\it Conformal anomaly of (2,0) tensor multiplet in six-dimensions
and AdS / CFT correspondence,}
JHEP 0002 (2000) 013, hep-th/0001041.

\bibitem{DDI} S. Deser, M.J. Duff and C.J. Isham,
{\it Nonlocal Conformal Anomalies,}
Nucl. Phys. {\bf B111} (1976) 45.

\bibitem{Duff77} M.J. Duff,
{\it Observations On Conformal Anomalies,}
Nucl. Phys. {\bf B125} (1977) 334.

\bibitem{DeserSchwimmer} S. Deser, and  A. Schwimmer,
{\it Geometric classification of conformal anomalies in arbitrary
dimensions,}
Phys. Lett. {\bf B309} (1993) 279, %% -284
%%   DOI: 10.1016/0370-2693(93)90934-A
hep-th/9302047.

\bibitem{FrTs-superconf} E.S. Fradkin, and A.A. Tseytlin,
{\it Asymptotic Freedom In Extended Conformal Supergravities,}
Phys. Lett. {\bf B110} (1982) 117; %% -122
%%  DOI: 10.1016/0370-2693(82)91018-8
{\it One Loop Beta Function in Conformal Supergravities,}
Nucl. Phys. {\bf B203} (1982) 157. %% -178
%%  DOI: 10.1016/0550-3213(82)90481-3

\bibitem{Paneitz} S. Paneitz,
{\it A Quartic Conformally Covariant Differential
Operator for Arbitrary Pseudo-Riemannian Manifolds,}
MIT preprint, 1983;
SIGMA {\bf 4} (2008) 036,
%% DOI: 	10.3842/SIGMA.2008.036
arXiv:0803.4331.

\bibitem{Hamada} K. Hamada,
{\it Integrability and Scheme Independence of
Even-Dimensional Quantum Geometry Effective Action,}
Prog. Theor. Phys. {\bf 105} (2001) 673, hep-th/0012053.

\bibitem{Wolfram}  Wolfram Research, {\it Mathematica,
Version 9.0}, Champaign, IL (2012).

\bibitem{decanini} Y. D\'{e}canini and A. Folacci,
{\it Irreducible forms for the metric variations of the action terms
of sixth-order gravity and approximated stress-energy tensor,}
Class. Quant. Grav. {\bf 24} (2007) 4777;
arXiv:0706.0691.

\bibitem{BD85} D.G. Boulware and S. Deser,
{\it String Generated Gravity Models,}
Phys. Rev. Lett. {\bf 55}, 2656 (1985),
DOI: 10.1103/PhysRevLett.55.2656.
%

\bibitem{capkim}
D.M.~Capper and D.~Kimber,
  {\it An Ambiguity in One Loop Quantum Gravity,}
J.\ Phys. {\bf A13} (1980) 3671.
%%  doi:10.1088/0305-4470/13/12/016

\bibitem{Tensors} I.L. Shapiro,
{\it Primer in Tensor Analysis and Relativity,}
(Springer, NY, to be published).
%
\bibitem{LPP13}
H. Lu, Y. Pang and C.N. Pope,
{\it Black Holes in Six-dimensional Conformal Gravity },
Phys. Rev. {\bf D87} (2013) 104013,
DOI: 10.1103/PhysRevD.87.104013.
%


\bibitem{Gauss} G.~de Berredo-Peixoto and I.L.~Shapiro,
{\it Conformal Quantum Gravity with the Gauss-Bonnet term},
Phys. Rev. {\bf D70} (2004) 044024,  hep-th/0307030;
\\
{\it Higher derivative quantum gravity with Gauss-Bonnet term,}
Phys. Rev. {\bf D71} (2005) 064005,
%%  doi:10.1103/PhysRevD.71.064005
hep-th/0412249.

\bibitem{star} A.A.~Starobinsky,
{\it A New type of isotropic cosmological models without singularity},
Phys. Lett. {\bf B91} (1980) 99.

\bibitem{star83} A.A.~Starobinsky,
{\it 	The perturbation spectrum evolving from a nonsingular initially
de-Sitter cosmology and the microwave background anisotropy,}
Sov. Astron. Lett. {\bf 9} (1983)  302.

\bibitem{StabInstab} T. de Paula Netto, A.M.~Pelinson, I.L.~Shapiro
and A.A.~Starobinsky,
{\it From stable to unstable anomaly-induced inflation,}
Eur. Phys. J. {\bf C76} (2016) 544,
%%  doi:10.1140/epjc/s10052-016-4390-4
arXiv:1509.08882.

\bibitem{anom2003}  M.~Asorey, E.V.~Gorbar and I.L.~Shapiro,
  {\it Universality and ambiguities of the conformal anomaly,}
  Class. Quant. Grav.  {\bf 21} (2003) 163,
%%  doi:10.1088/0264-9381/21/1/011
hep-th/0307187.

\end{thebibliography}
\end{document}